\documentclass[fleqn,10pt]{wlscirep}
\usepackage[utf8]{inputenc}
\usepackage[T1]{fontenc}
\usepackage{latexsym,epsfig,epstopdf}
\usepackage{amssymb,amsmath,amsthm}
\usepackage{xcolor}
\usepackage{color}
\usepackage{ulem}
\usepackage{gensymb}

\title{Numerical tests of magnetoreception models assisted with behavioral experiments on American cockroaches}

\author[1,*]{Kai Sheng Lee}
\author[1,2]{Rainer Dumke}
\author[1,3]{Tomasz Paterek}

\affil[1]{School of Physical and Mathematical Sciences, Nanyang Technological University, 637371 Singapore, Singapore}
\affil[2]{Centre for Quantum Technologies, National University of Singapore, 117543 Singapore, Singapore}
\affil[3]{Institute of Theoretical Physics and Astrophysics, Faculty of Mathematics, Physics, and Informatics, University of Gda\'nsk, 80-308 Gda\'nsk, Poland}
\affil[*]{Correspondence and requests for materials should be addressed to K.S. (email: kaisheng001@e.ntu.edu.sg)}

\begin{abstract}
Many animals display sensitivity to external magnetic field, but only in the simplest organisms the sensing mechanism is understood.
Here we report on behavioural experiments where American cockroaches (Periplaneta americana) were subjected to rotating external magnetic fields with a period of 10 minutes.
The insects show increased activity when placed in a rotating Earth-strength field, whereas this effect is diminished in ten times stronger rotating field.
We analyse established models of magnetoreception, the magnetite model and the radical pair model, in light of this adaptation result.
Our findings show that the magnetite model is excluded and the radical pair model requires strong additional assumptions to be compatible with the data.
\end{abstract}
\begin{document}

\flushbottom
\maketitle
%
%
\thispagestyle{empty}

\section*{Introduction}

The Earth's magnetic field has existed for at least $3.5$ billion years, the result of an electrically conducting fluid core \cite{Weiss2002}. Most of life has evolved in the presence of the Earth's field and it is unsurprising that organisms have developed adaptations taking advantage of it. For instance, magnetotactic bacteria are observed to passively align with magnetic field lines, a phenomena coined as magnetotaxis \cite{Blakemore1975}. 
This is achieved via magnetic crystals encased in membranes, and a small enough mass for the magnetic torque to steer the organism. 
For larger animals, an added layer of complexity in translating magnetic information to biologically useful neuronal signals is required. 
We refer to this process of sensing and translating as magnetoreception or magnetic sensing (we do not require that it acts as a compass, i.e. provides directional information).

A plethora of species across the animal phyla, from insects like planthoppers \cite{Pan2016} or honeybees \cite{Kirschvink1981bee,KIRSCHVINK1991,Kirschvink1363}, fish like yellowfin tuna \cite{WALKER1984} or sockeye salmon \cite{Mann35}, migratory birds like homing pigeons \cite{Tian2006} and  mammals like bats \cite{Holland2008,Tian2010}, have been observed to exhibit the ability of magnetoreception. A fuller compilation of known species can be found in \cite{Wiltschko2006}. 
However, beyond magnetotactic bacteria, the mechanism behind magnetoreception is not known. The usual candidate explanations include the magnetite model~\cite{Kirschvink1981} and the radical pair model~\cite{Schulten1978,Hore2016,Ritz2000}. The former involves the presence of ferromagnetic deposits that act like tiny compasses, while the latter involves chemical reactions with distinct products that can be modulated by an external magnetic field.
It is suggested that both methods might be present in a single animal~\cite{Munro1997}.
Alternative ideas exist, for example, many experimental findings are compatible with the model based on magneto-elastic properties of cells~\cite{Krichen2017}.
Clearly, more data is required to narrow down theoretical possibilities and ultimately localise relevant receptor organs and sensory pathways~\cite{Johnsen2005}.

A major practical significance of magnetoreception is not only the ability to sense the Earth field of approximately 0.5 G (Gauss) but also sensitivity to time-varying fields with amplitudes in the range $10-100$ $\mu$G, observed in European robins \cite{Thalau2004} and American cockroaches~\cite{Vacha2009}. 
This is achieved by biological sensors at room temperature and compact sizes, and once understood will lead to robust man-made magnetic sensors.

Here we report on experiments with American cockroaches (Periplaneta americana) that confirm their magnetoreception and show adaptation of the sensory mechanism to the Earth's magnetic field.
We then use these results to put constraints on the magnetite and radical pair models.
Cockroaches are good candidates for studies on magnetic sensing for several reasons.
Their genome has been completely sequenced~\cite{Li2018}, opening the way towards controlled and focused genetic studies related to magnetoreception.
The size of cockroaches makes them easy to handle and translates to compact table top experiments. Cockroaches were shown to be magnetisable, with very long magnetisation decay ranging from about an hour in a living insect to about two days in a dead one~\cite{Kong2018}.
Finally, cockroaches have already been observed to react to changes in external magnetic fields~\cite{Vacha2006,Vacha2009,vacha2010}, and it is known that the sensing involves the protein cryptochrome \cite{Bazalova2016}. 
All these behavioural experiments have been conducted in the group of Vacha, and the added value of the present work is an independent confirmation of magnetoreception in Periplaneta.

In summary, we video recorded cockroaches in various magnetic fields and utilised tracking software to extract the time when they were active (rotating or translating).
Cockroaches are found to be more active in a periodically rotating Earth-strength magnetic field. When instead they were faced with a rotating magnetic field of $5$ G, ten times the strength of the geomagnetic field, cockroach activity is comparable to the value observed in control experiments where field rotations were absent. 
We simulate leading explanations of magnetoreception and conclude that the magnetite model cannot explain our data, whereas the radical-pair model with additional assumptions is not excluded, but implausible.

\section*{Results}

We first describe our methodology, inspired by that of Vacha and extending Refs.~\cite{Vacha2006,Vacha2009,vacha2010,Bazalova2016}, and then present obtained results.

\begin{figure}[!t]
	\centering
	\includegraphics[width=14cm]{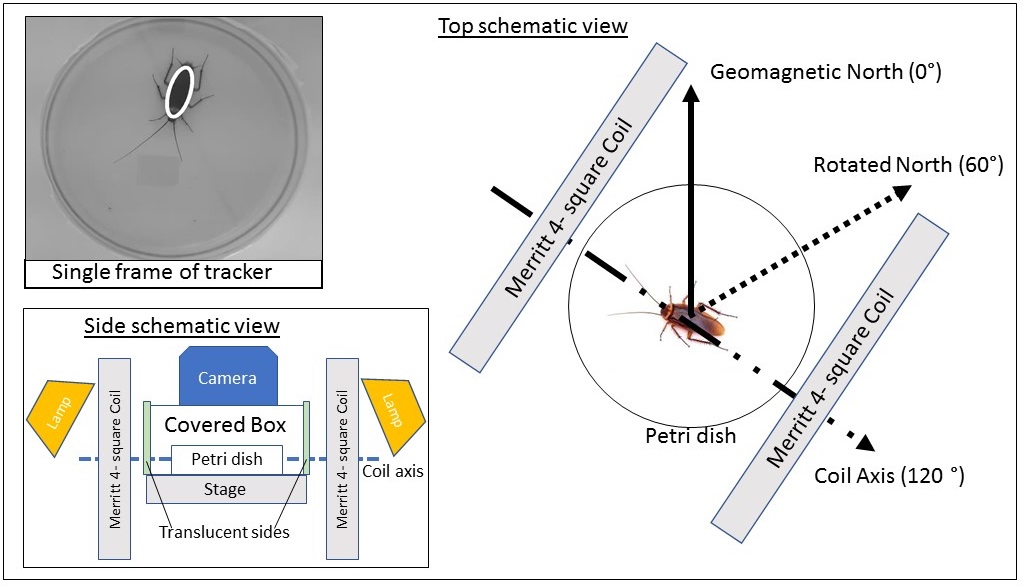}
	\caption{Schematic of experimental setup. Cockroaches were placed in Petri dishes and levelled inside the Merritt coil used to rotate the magnetic north by 60$\pm 5\degree$ clockwise (preserving inclination angle). The coil alternates between being switched on and off every 5 minutes, from 6 am to 6 pm. A camera records cockroach motion from the top of a covered box prepared to minimise visual cues from surrounding (bottom left inset). Tracking software recognises the insect (top left inset showing ellipse fitted to the cockroach) and stores its location and angle in the horizontal plane. Our figure of merit is the activity time defined as the time a cockroach spends translating or rotating.}
	\label{fig:cockroachcoil}
\end{figure}

The schematic of our setup is depicted in Fig.~\ref{fig:cockroachcoil}.
Experimental procedure and data processing is described in detail in the Methods section. In essence, the cockroaches were video recorded in an isolated room by a fully automated apparatus.
A camera was attached on top of the box that contained a single cockroach in a Petri dish at a time. Visual cues from the surrounding were minimised.
The box was installed inside the Merritt coil,
a configuration of four square loops of wires that produced the uniform magnetic field across the area accessible to the cockroach.

Randomised permutations of experimental conditions and cockroach specimen were made, with the constraint that the same cockroach is not re-used in consecutive experimental runs. Three experimental conditions were studied: one control and two different classes of tests. In the control runs (N=29), the coils were switched off at all times. In the first class of tests, with Earth-strength field of $0.5$ G (N=29), the coil was arranged such that, when turned on, its field rotates the magnetic north by 60$\pm 5 ^\circ$ clockwise in the horizontal plane. The magnitude and inclination angle of the rotated field was unchanged with respect to the geomagnetic field. The coil was then periodically switched on and off every five minutes, from 6 am to 6 pm. In the second class of tests, with the field of $5$ G (N=16), a different Merritt coil was used to produce a 60$\pm 5^\circ$ rotated field which is $10$ times stronger than the Earth's magnetic field. The same switching profile was used as in the first class of tests. 

\begin{figure}[!t]
	\centering
	\includegraphics[scale=0.5]{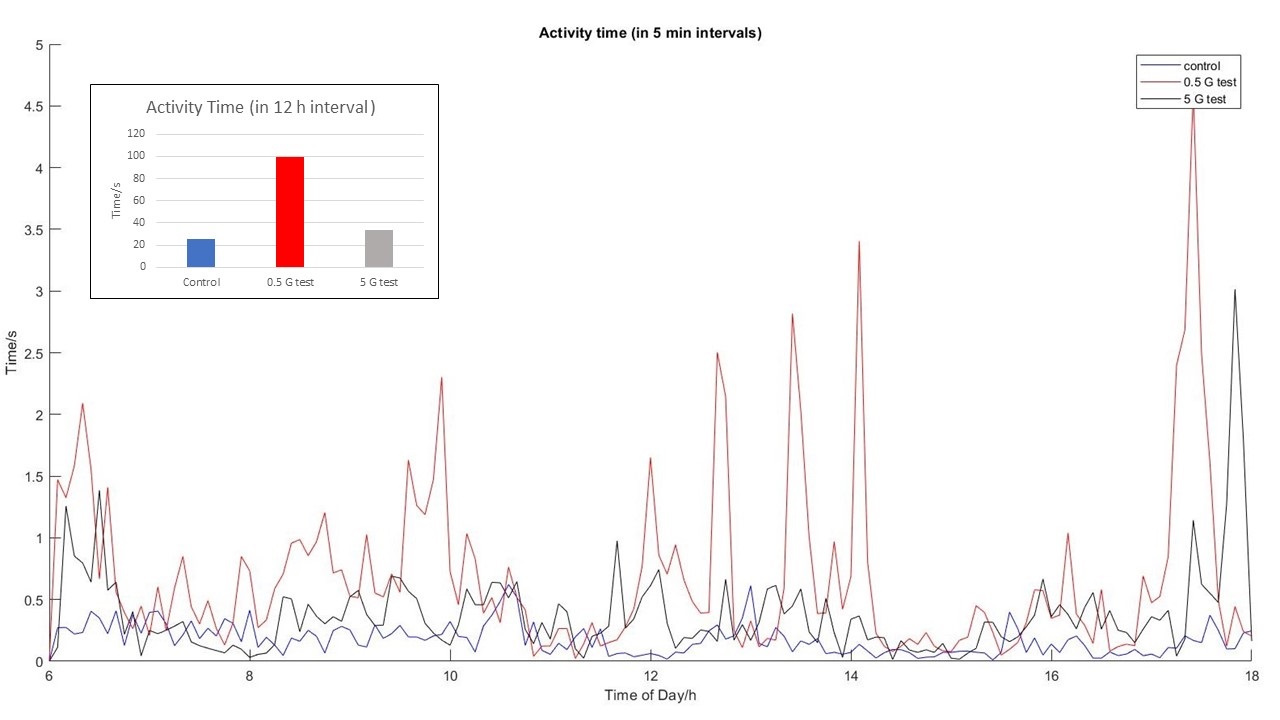}
	\caption{Activity of the American cockroach. The main plot shows the mean activity time, i.e. the time an average cockroach was translating or rotating. The curves are obtained by connecting points showing the mean activity in the preceding $5$ minute time interval. The inset shows the mean activity accumulated over the time span of the whole experiment ($12$ hours).
	}
	\label{fig:cockroachexpt}
\end{figure}

From the obtained videos, cockroach position and angle in the horizontal plane was extracted for every frame and used to compute the activity time, defined as the time the cockroach translates or rotates.
The results of all the experiments are summarised in Fig.~\ref{fig:cockroachexpt}.

\section*{Discussion}

The results of the controls show that cockroaches tend to be stationary over the day periods (note that the activity time in Fig.~\ref{fig:cockroachexpt} is in seconds).
However, the activity is increased when the rotating field has the strength of the geomagnetic field. In a stronger $5$ G field this effect is diminished and comparable with the controls.
This confirms sensitivity of the American cockroach to the direction of external magnetic field observed in group of Vacha and indicates that the sensing mechanism is evolutionarily tuned to the Earth's field.

We now discuss two main candidate models of magnetoreception in the light of the obtained results.

\subsection*{Magnetite model}

The magnetite model supposes that magnetic materials inside the animal rotate as little compasses in an external magnetic field. These rotations are transduced to the nervous system and interpreted.
The hypothesis gained popularity with the discovery that certain bacteria are capable of precipitating magnetic grains (either magnetite Fe$_3$O$_4$~\cite{Blakemore1975,frankel1979,Balkwill1980} or greigite Fe$_3$S$_4$ \cite{Heywood1990} or both~\cite{Bazylinski1995}) via specialised organs.
Other animals, including humans, can also mineralise magnetic particles~\cite{Kirschvink1992}.
One also verifies plausibility of this model by noting that magnetic energy of the single domain grain of magnetite with radius $50$ nm placed in the Earth's magnetic field is almost $10$ times larger than the thermal energy at room temperature.
Furthermore, the cockroaches were observed to acquire magnetic moment in the presence of a strong magnetic field, confirming presence of magnetic materials in their bodies~\cite{Kong2018}.
Due to observed long demagnetisation times Kong \textit{et al.} estimated that a magnetite particle of radius $R = 50$ nm, saturation magnetisation $M_s=3 \times 10^5$ A / m and mass density $\rho = 4.049 \times 10^3$ kg / m$^{3}$ has to rotate in medium with high viscosity $\eta = 10^5$ Pa s~\cite{Kong2018}.
We now demonstrate that such magnetic grains and their surroundings are disqualified by our data.

We conducted simulations of a set of $36$ spherical magnetic particles with parameters given above.
We monitored the motion in the horizontal plane and initialise the magnetic grains to have magnetisation axes with angular orientations uniformly distributed between $0$ and $2 \pi$, i.e. each particle (magnetic moment) is initially rotated by $10$ degrees from its neighbours. This has the macroscopic effect of cancelling out the magnetisation contributions from each individual grain, agreeing with the experimental observation that cockroaches have no residual magnetisation~\cite{Kong2018}, 

In the presence of an external magnetic field, the magnetites gradually align toward the field. The rotational motion of the $i$th particle is described by Newton's law:
\begin{equation}
I \ddot{\theta_i} = -f \dot{\theta_i}-\mu B \sin \theta_i,
\end{equation}
where $\theta_i$ is the angle between the external field and the $i$th magnetic moment, $I = \frac{2}{5}\rho V R^2$ is its moment of inertia, and $f = 8\pi \eta R^3$ the rotational friction coefficient.
Note that we have ignored the thermal torque which will additionally misalign the particles and reduce magnetisation of the whole set.
We have simulated the dynamics of the particles for $12$ hours changing the direction of the external field every $5$ minutes as in the experiment. 
Our figure of merit was the ``alignment'' defined as normalised total magnetisation:
\begin{equation}
M = | \sum_{i = 1}^n \vec m_i | / n
\label{EQ_M}
\end{equation}
where $n$ is the total number of particles and $\vec m_i$ is the magnetic moment of the $i$th grain.
$M$ is equal to $0$ when the individual moments exactly cancel out and it is given by $1$ when all the magnetic moments are parallel. 

\begin{figure}[!t]
	\centering
	\includegraphics[scale=0.55]{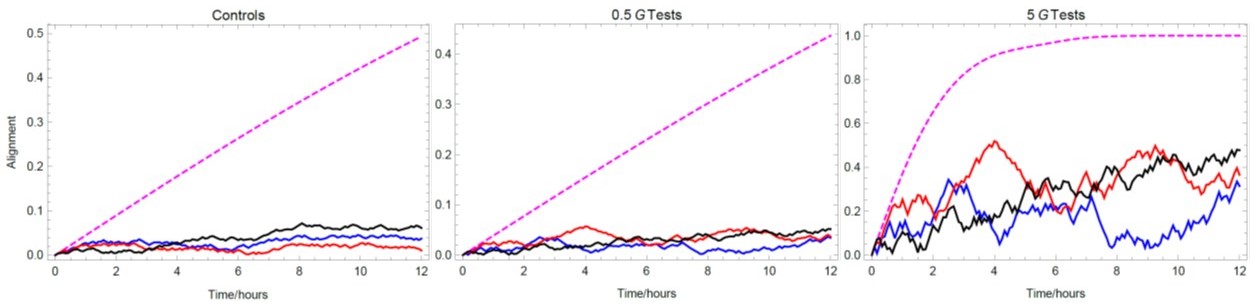}
	\caption{Alignment (normalised magnetisation) in the magnetite model. The dashed line shows the alignment, see Eq.~(\ref{EQ_M}), for a stationary cockroach. The solid colorful lines take into account the actual motion of three typical insects.}
	\label{fig:totalmag}
\end{figure}

The results of our simulations are presented in Fig.~\ref{fig:totalmag} and in the Supplementary Information (SI).
It turns out that in the presence of an Earth-strength magnetic field, a highly viscous environment leads to the alignment of only about $40 \%$ of all the magnetites after $12$ hours. In the $5$ G field, the alignment is complete. 
This is opposite to experimental findings, where cockroaches react to rotations of the Earth-strength field but not to rotations in the stronger field.
Simulations also show that one can trade the strength of external field with time.
The shape of the alignment curves in $5$ G field and in the longer simulations with $0.5$ G field is similar.
In SI we show that independently of the strength of the rotating field ($0.5$ G or $5$ G) the particles (partially) align to the angle of $30$ degrees, in between the geomagnetic and rotated fields.

Additionally, we ask if the motion of the cockroach is an attempt to maintain a certain preferred alignment. 
For this, let us denote the measured angle of the insect in the $j$th frame by $\theta_j^{\textrm{exp}}$.
We now modify the angle of the magnetic grain at times $t_j = j \Delta t$, where $\Delta t$ is the time duration between the frames, by adding to them the cockroach motion, i.e. $\theta_j \rightarrow \theta_j +\theta_j^{\textrm{exp}}$.
The simulations show that the motion does not lead to improved alignment. 
In fact, in the Earth-strength field, the motion inhibits $M$ which stays below $0.05$ level.
Note that this holds for both control and test experiments, again in disagreement with experimental finding that cockroaches behave differently in these two cases.


\subsection*{Radical pair model}

The radical-pair mechanism was first conceived in 1969 by Closs, Kaptein and Oosterhoff (CKO model) \cite{Closs1969,Kaptein1969} as an explanation of chemically induced dynamic nuclear polarization, which has ever since been an important technique in NMR spectroscopy \cite{Buchachenko2001,Goez2012}. It was later proposed by Schulten \textit{et al.}~\cite{Schulten1978} to be involved in animal magnetoreception. 

The Cryptochrome / Photolyase flavoprotein family are thought to be relevant to light-sensitive biological compasses \cite{Ritz2000}, hinting at the radical pair model. 
We base our discussion on the work by Solov'yov \textit{et al.}~\cite{Solovyov2007} who described creation of radical pairs and their dynamics in Cryptochrome-1 (Cry-1) of the plant Arabidopsis thaliana. While Cry-1 was absent in Periplaneta americana, Cryptochrome-2 was found to be present and also necessary for magnetic sensitivity \cite{Bazalova2016}. The parameters below are taken from experiments on Photolyase in E. coli (hyperfine axes)\cite{Cintolesi2003} and Cry-1 in Arabidopsis thaliana (transition rates)\cite{Byrdin2004,Aubert2000}. This is justified given that the radical pair mechanism is believed to be contained within the highly conserved FAD molecular domain \cite{Mei2015}, common across the Cryptochrome / Photolyase family.

\begin{figure}
	\centering
	\includegraphics[scale=0.9]{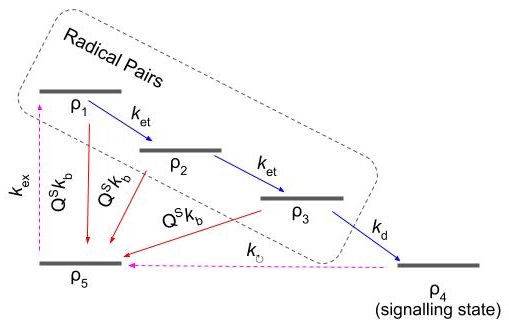}
	\caption{Effective model of the cryptochrome radical pair mechanism. 
	The first radical pair is created in the state $\rho_1$ by an excitation of the precursor molecule $\rho_5$. Electron hopping gives rise to other pairs, $\rho_2$ and $\rho_3$. Their spin dynamics depends on the external magnetic field and they can recombine to the precursor only if they are in the singlet state. This selective recombination is represented by the singlet projectors $Q_s$ in front of the back-transfer rates, $k_b$, and modifies the yield of the signalling state $\rho_4$, perceived by the animal. Variants of the model with solid transitions have been studied previously, see e.g.~\cite{Solovyov2007,Chia2016_1,Chia2016_2}.
	We also consider the model with additional dashed arrows, which close the dynamics of the whole system.}
	\label{fig:effectivemodel}
\end{figure}

The radical-pair reaction pathway can be compressed to essential steps given in Fig.~\ref{fig:effectivemodel}.
The model begins with the creation of the pair of electrons (each on a different radical molecule) in the state $\rho_1$, being initially the singlet state.
Due to the proximity of electron-accepting molecules, one of the electrons may hop around, giving rise to a chain of radical pairs $\rho_1 \rightarrow \rho_2 \rightarrow \rho_3$ connected with electron transfer rate $k_{\mathrm{et}}$.
This rate has been measured to be on the order of $10^8$ Hz~\cite{Solovyov2007}.
There is also a possibility of reverse electron transfer,
but its rate is estimated to be two orders of magnitude smaller and therefore we ignore it.
The whole system evolves in the external magnetic field $\vec B$ and additionally both electrons are coupled to their nearby nuclei, so that the total Hamiltonian is the sum of the following terms:
\begin{equation}
\label{eq:hamil}
H_j= 2 \mu_B \vec{B} \cdot \vec{S}_j + \mu_B \sum_i \vec I_i \cdot (\overleftrightarrow{A}_{ij}^{\textrm{iso}} + \overleftrightarrow{A}_{ij}^{\textrm{aniso}}) \cdot \vec{S}_j,
\end{equation}
where the subscript $i=1,2$ refers to the nuclei, 
$j=1,2$ to the electrons, and $\mu_B$ is the Bohr magneton.
The first term is the Zeeman interaction of the electron spin $\vec S_j$ with the external magnetic field.
The second terms is the hyperfine interaction with nuclear spin $\vec I_i$ that couples to the electronic spin with diagonal hyperfine tensor $\overleftrightarrow{A}_{ij}^{\textrm{iso}}$ and anisotropic tensor $\overleftrightarrow{A}_{ij}^{\textrm{aniso}}$.
Both of these have been measured~\cite{Kay1999,Weber2001,Cintolesi2003} and we summarise the values used in our simulations in SI.
Due to this interaction, the pair coherently oscillates between singlet and triplet states and can selectively recombine to the precursor of the radical pairs $\rho_5$, with the estimated rate $k_b = 10^7$ Hz, only if the pair is in the singlet state.
Finally, the last pair in the chain, $\rho_3$, can decay to the signalling state $\rho_4$, with the measured rate $k_d = 3.3 \times 10^6$ Hz. The amount of $\rho_4$ is assumed to be the chemical perceived by the animal.
All the transfers are incoherent processes and are described with the help of the rate equations. The final figure of merit is the yield of the signalling state $\Phi$ being the probability of finding the system in $\rho_4$ after a long time.

\begin{figure*}
	\centering
	\includegraphics[scale=0.35]{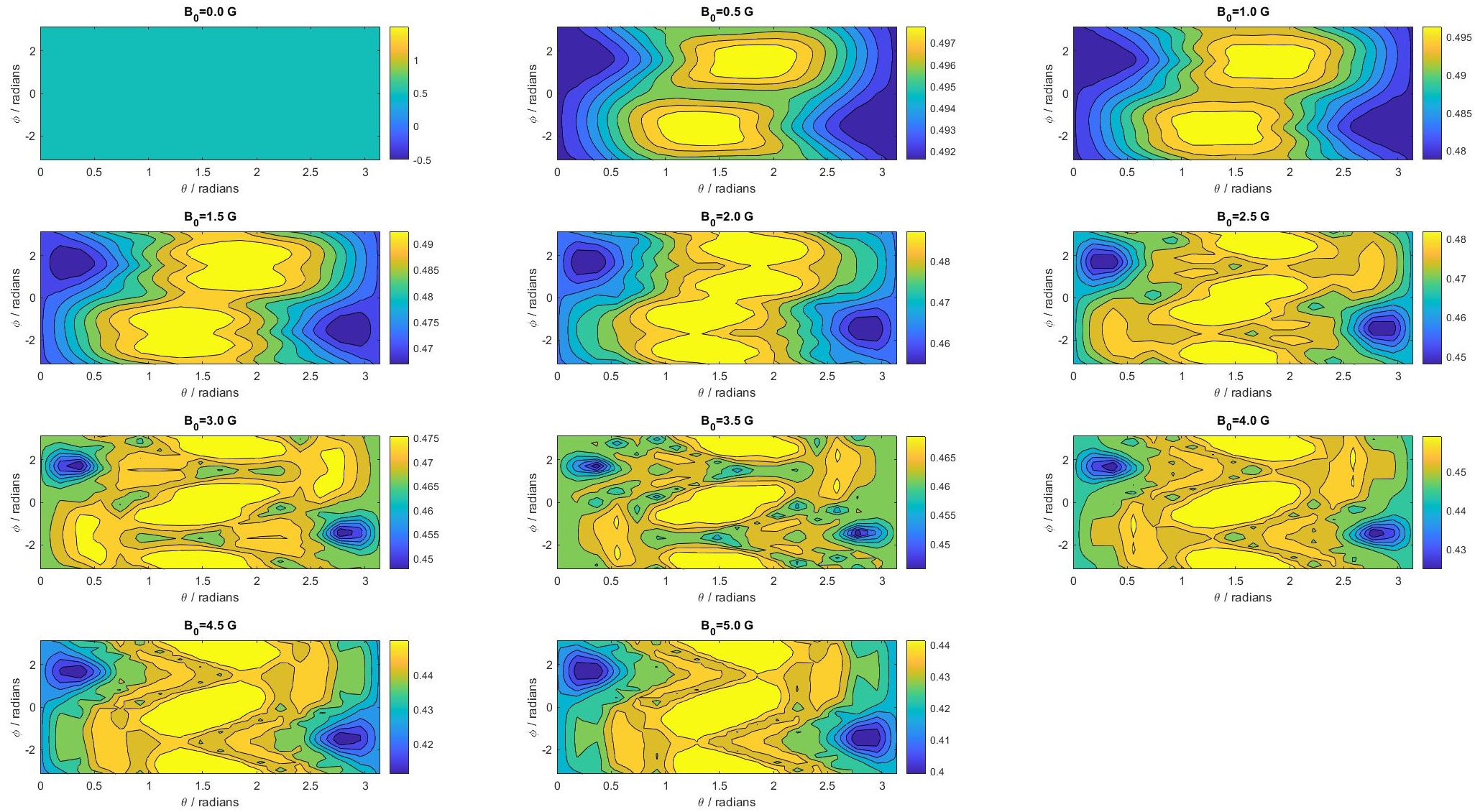}
	\caption{Signalling yield as a function of external magnetic field in the radical pair model with solid arrows in Fig.~\ref{fig:effectivemodel}.
		The field is parameterised in spherical coordinates $\vec{B}=(B_0 \sin \theta \cos\phi, B_0 \sin\theta \sin\phi, B_0 \cos\theta)$.
		The hyperfine tensors used in the simulation were measured using EPR and ENDOR techniques~\cite{Kay1999,Cintolesi2003}.
	The more complex structures visible for stronger external fields are the effect of comparable hyperfine interactions (in the range $4-16$ G) and external fields.}
	\label{fig:realyield}
\end{figure*}

In SI we describe validation of our numerics with analytical estimates and compare our simulations to those of Ref.~\cite{Solovyov2007}.
Fig.~\ref{fig:realyield} shows $\Phi$ as a function of spherical polar and azimuthal angles of the external magnetic field of various magnitudes.
There is a variation of the yield with the direction of the external field for all magnitudes.
In particular, for the Earth-strength field of $0.5$ G, the contrast $\Phi_{\max} - \Phi_{\min}$, is around $0.005$ and yield is about $0.49$, whereas in the $5$ G field, we have a contrast of $0.04$ and a yield smaller than $0.44$.
In principle, the result of our experiment could be explained within this model by adding an assumption that the sensing requires the signalling yield above a threshold value of about $0.45$, and that changes in the yield on the order of one in a thousand are perceivable by the animal.

\begin{figure*}[!b]
	\centering
	\includegraphics[scale=0.35]{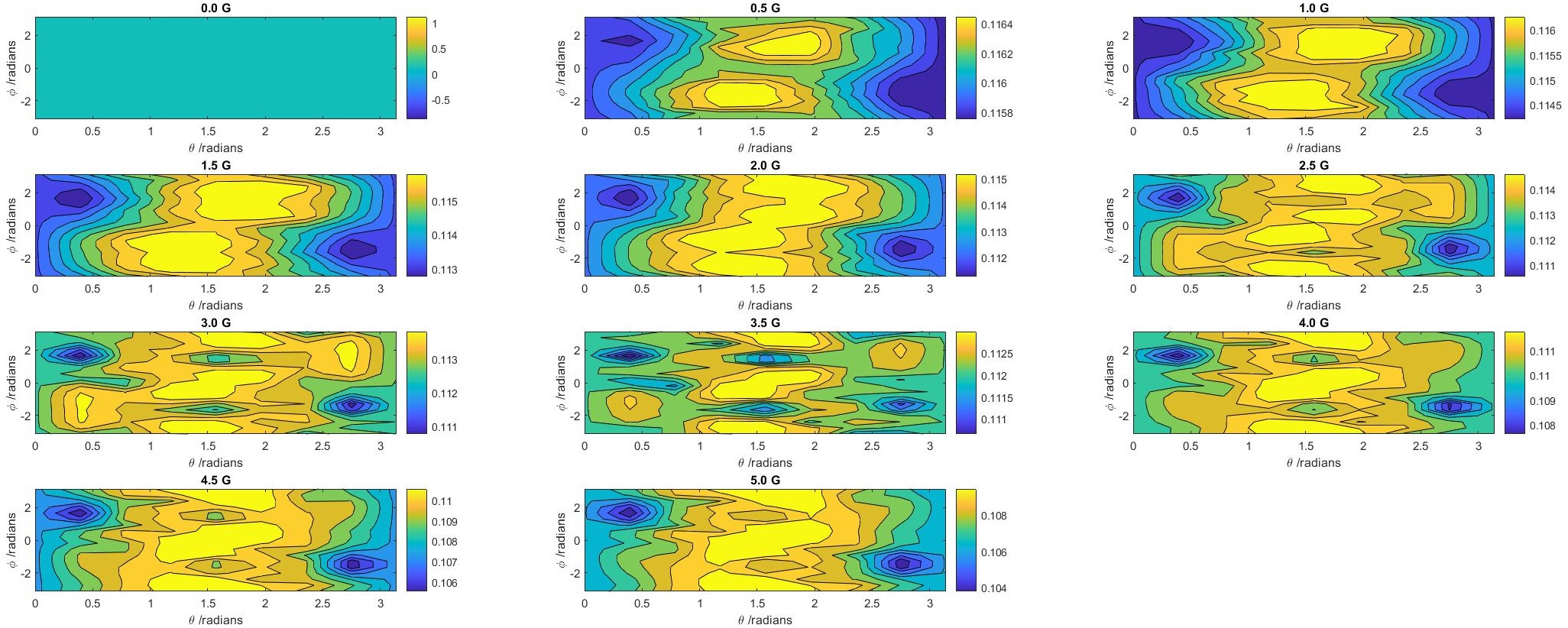}
	\caption{Signalling yield as a function of external magnetic field in the radical pair model with closed dynamics, i.e. with additional dashed magenta arrows in Fig.~\ref{fig:effectivemodel}. The model preserve compass features, but the yield and variations are diminished.}
	\label{fig:10^7}
\end{figure*}

The model that was just described begins with the formation of radical pair and ends with the signalling state as a sink.
Many natural processes are closed and we now ask about the possibility of a self-sustaining biocompass that can recycle the signalling state.
For this, we add to the presented radical-pair model two transitions (denoted with dashed magenta lines in Fig.~\ref{fig:effectivemodel}) that close the loop of the process.
The transition $\rho_5 \to \rho_1$ occurs with the rate $k_{\textrm{ex}}$ and captures the rate of creation of the first radical pair, typically as a result of illumination with sunlight.
The transition $\rho_4 \to \rho_5$ occurs with the rate $k_{\circlearrowright}$ and describes conversion of the signalling chemical to the initial precursor molecule.
Numerical simulations show that the compass retains its functionality only if both of these new rates are similar.
If $k_{\circlearrowright} \gg k_{\textrm{ex}}$ the system accumulates in the precursor state $\rho_5$ and if $k_{\circlearrowright} \ll k_{\textrm{ex}}$ the steady state is the signalling state $\rho_4$ independently of the external magnetic field.
Fig.~\ref{fig:10^7} shows the results for exemplary set of $k_{\textrm{ex}} = 5 \times 10^6$ and $k_{\circlearrowright} = 10^7$.
The variations of the yield are qualitatively similar to those in Fig.~\ref{fig:realyield} but quantitatively the yield is about ten times smaller and its variations are also correspondingly smaller.
This shows that closing the dynamics has important consequences and makes the radical pair hypothesis rather implausible.

\section*{Conclusions}

In conclusion, we conducted behavioural experiments on American cockroaches (Periplaneta americana) that confirm their sensitivity to directional changes of Earth-strength magnetic fields of $0.5$ G.
The sensitivity is revealed by increased activity of insects during daytime.
Furthermore, the data from experiments in $5$ G rotating fields show diminished activity, indicating adaptation of the sensing mechanism to the Earth's magnetic field. A similar observation has been made for European robins exposed to fields $2$ times stronger than the Earth's field \cite{Wiltschko2006_2}. We performed numerical analysis of the usual theoretical candidates of magnetoreception, the magnetite and radical-pair models, in light of the obtained results.
Our analyses exclude the magnetite model with parameters established in previous experiments on cockroaches.
They also show that the radical-pair model, with the parameters presently available from the experiments on Photolyase and Cry-1, can only explain the sensitivity to the Earth-strength field and its diminishing effect in the stronger field,
if we assume that contrast in the chemical yield on the order of one in a thousand is perceived by the animal, and additionally that this only happens if the yield is above a certain threshold (attained in $0.5$ G but not in $5$ G field).

\section*{Methods}

\subsection*{Experimental procedure}

The experiments were conducted in Singapore.
Adult female and male cockroaches were kept in separate transparent insectaria with unlimited water, a diet consisting of cat food pallets and photo-period of $12$ light (6am to 6pm) : $12$ dark (6pm to 6am) hours.
A day before the experiment, at 6 pm, a single insectarium was placed in a $4^{\circ}$ C environment in order to immobilise the insect, which was then moved to a Petri dish ($15$ cm diameter) with circumference covered with white slip.
To minimise visual cues, the Petri dish was placed in a box with the inside surfaces covered by white paper, except two opposite facing sides with white translucent films to allow external lights for uniform illumination inside the box. A small aperture was made on the overhead for the camera configured with a capture rate of $30$ frames per second.
The box was installed on a stage levelled in the middle zone (where the generated magnetic field is most uniform) of a Merritt four square coil~\cite{Merritt1983}.

The cockroach was left overnight in an isolated room in order to acclimatise with the new environment. The experiment automatically begins at $6$ am the day after. 
In the first class of test runs, the magnetic field was alternating between the natural geomagnetic field in Singapore and the $60\pm5 \degree$ clockwise-rotated field.
This rotation was realised by switching on the Merritt coil (side dimension $1.2$ m), arranged at $120\pm5^\circ$ from the geomagnetic north. This modifies the horizontal component of the field while keeping the magnitude and inclination angle of the Earth's field.
In the second class of test experiments we used a smaller Merritt coil (side dimension $0.3$ m) arranged at suitable angle in order to generate the field rotated by $60\pm5^\circ$ from the geomagnetic direction (in the horizontal plane) with the strength of $5$ G, i.e. about $10$ times stronger than the Earth's magnetic field. The remaining procedures were the same.
The same procedures were also followed in the control runs, except that the coils were not switched on at all.

\subsection*{Data analysis}

The experiments output video recordings of cockroach motion in the Petri dishes.
In order to obtain numerical parameters easily comparable between tests and control runs the videos were processed as follows.
The self-written tracking software identified the cockroach in every frame, fitted an ellipse to it, and stored in a text file timestamp, coordinates of the center of the ellipse and angle of the main axis.
From these values we computed activity time by summing up time intervals between the frames ($33.3$ ms) in which the center location changed by more than $3$ mm or the angle changed by more than $8$ degrees.

\section*{Appendix A: Alignment in the magnetite model}
\begin{figure}[!h]
	\centering
	\includegraphics[scale=0.6]{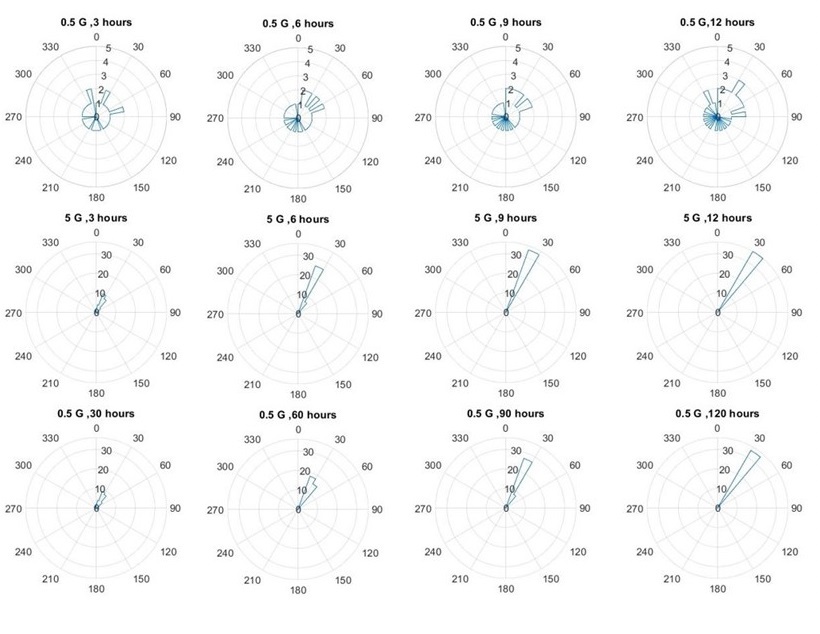}
	\caption{Alignment in the magnetite model. The angular plots show histograms for the orientation angles of the $36$ magnetic particles at different times and different strengths of external magnetic fields as labelled. Initially the particles were distributed uniformly over the circle. Note that the angle of final alignment is $30^\circ$. Note also that time for alignment is almost interchangeable with magnetic field strength, as seen from similarities between the middle and bottom histograms.}
	\label{fig:directionalplots}
\end{figure}

Here we show that the magnetite model predicts alignment to the angle between the geomagnetic and rotated fields in both studied classes of test experiments. The results of the simulations are shown in Fig.~\ref{fig:directionalplots}.

\section*{Appendix B: Hyperfine tensors}
\label{APP_VALUES}

In our simulations of the radical pair model we used the hyperfine tensors as shown in Tab. \ref{tab:hyperfine}, which have been measured using EPR and ENDOR techniques (see compilation in Ref.~\cite{Cintolesi2003}).
Note that these values are different from those used in Ref.~\cite{Solovyov2007}.

\begin{table}
\centering
\begin{tabular}{llllll}
\hline \hline
\multicolumn{5}{c}{\quad First electron interacts with one nucleus \quad }\\
\hline
$A^{\textrm{iso}}_{11}$ [G] \quad & $A^{\textrm{aniso}}_{11}$ [G] \quad & \multicolumn{3}{c}{hyperfine axes}\\
\hline
3.93 & -4.98 & 0.4380 & 0.8655 & -0.2432   \\
    & -4.92 & 0.8981 & -0.4097 & 0.1595    \\
    & 9.89 & -0.0384 & 0.2883 & 0.9568    \\
\hline \hline
\multicolumn{6}{c}{\quad Second electron interacts with two nuclei \quad } \\
\hline
$A^{\textrm{iso}}_{12}$ [G] & $A^{\textrm{aniso}}_{12}$ [G] & \multicolumn{3}{c}{hyperfine axes}\\
\hline
 13.6 & 0 & 1 & 0 & 0   \\
    & 0 & 0 & 1 & 0   \\
    & 0 & 0 & 0 & 1    \\
    \hline
$A^{\textrm{iso}}_{22}$ [G] & $A^{\textrm{aniso}}_{22}$ [G] & \multicolumn{3}{c}{hyperfine axes}\\
\hline
 -4 & -0.23 & -0.984 & 0.180 & 0   \\
    & 0.35 & 0.180 & 0.984 & 0   \\
    & -0.12 & 0 & 0 & 1    \\
\hline
\end{tabular}
\caption{Hyperfine tensors used in our simulations of the radical pair model.}
\label{tab:hyperfine}
\end{table}


\section*{Appendix C: Validation of numerics}
\label{APP_SOLOV}

We describe here in more detail how we have verified accuracy of our numerics as certain obtained results are different from those in the literature.
In particular, Ref.~\cite{Solovyov2007} computes the variation in time of the probability that the radical pair $\rho_3$ is in the singlet / triplet state. 
Both are found to be at most $0.05$ within the first $500$ ns of evolution (Fig. 10 of that reference).
We now give analytical estimates of these probabilities which turn out to be an order of magnitude higher and provide the following intuitive explanation why higher values should be expected.
Recall that the electron transfer rate is an order of magnitude faster than the recombination rates and two orders of magnitude faster than the decay rate to the signalling state.
Therefore, all the dynamics in the chain $\rho_1 \to \rho_2 \to \rho_3$ is fast and one expects non-negligible portion of pairs in the state $\rho_3$.

\begin{figure}[h!]
	\centering
	\includegraphics[scale=0.3]{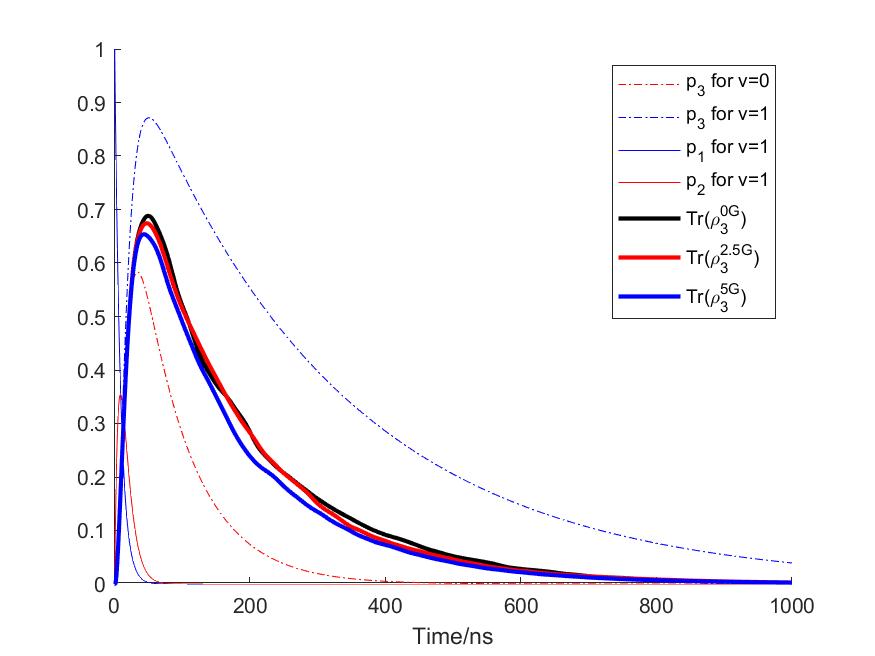}
	\caption{Population dynamics of the radical pairs. 
		The solid curves are computed with the rates and the hyperfine tensors as in Ref.~\cite{Solovyov2007}. The population of the first two radical pairs is transferred to the third pair within $50$ ns. The population of the third pair is plotted in thick lines for various strengths of external magnetic field. All of them lie within the obtained analytical bounds shown by dashed-dotted lines.
		}
	\label{fig:bounds}
\end{figure}

In order to place the analytical bounds on the probability that the system is in $\rho_3$ (independently of whether it is the singlet or triplet state) we note that setting the recombination rate to zero gives the upper bound to the population in $\rho_3$, whereas allowing recombination independent of the spin state provides the lower bound on the population.
Since in both cases the system looses spin dependence it is governed by the following set of rate equations, which can be read out from the reaction scheme in Fig. 4 of the main text:
\begin{align}
\dot{p}_1 &=(-k_{\textrm{et}} - v \, k_b) \, p_1, \nonumber \\
\dot{p}_2 &=(-k_{\textrm{et}} -v\, k_b) \, p_2 + k_{\textrm{et}} \, p_1, \nonumber \\
\dot{p}_3 &=(-k_{d} -v\, k_b) \, p_3 + k_{\textrm{et}} \, p_2,
\end{align}
where $p_j$ is the population in the $j$th radical pair and $v = 0,1$ turns on and off the possibility of recombination. 
This set of equations admits analytical solution:
\begin{align}
p_1 & =\exp[(-v\, k_b - k_{\textrm{et}})t],\\
p_2 &= k_{\textrm{et}} \, t \, p_1, \nonumber \\
p_3 &= \frac{k^2_{\textrm{et}}}{(k_{\textrm{et}}-k_d)^2} \left[\exp\left((-v\, k_b-k_d)t \right)- p_1 \right] - \frac{k_{\textrm{et}}}{k_{\textrm{et}} - k_d} \, p_2. \nonumber
\end{align}
The corresponding limiting curves (for $v=0$ and $v=1$) are plotted in Fig.~\ref{fig:bounds} (dashed-dotted lines). 
The plot confirms that the populations $p_1$ and $p_2$ quickly decay to zero and shows that the number of radical pairs $\rho_3$ lies well within the obtained boundaries.
Note that the results in Ref.~\cite{Solovyov2007} are below the analytical lower bound obtained here.

\section*{Acknowledgements}
We would like to thank Zhang Wei for preparing the first version of the magnetic coils, Anthony Tan and Ng Hong Kuan for writing and testing the tracking software, Agnieszka G\'orecka and Herbert Crepaz for assistance in early experiments, and Dagomir Kaszlikowski for quantum biology spirit.
This work is supported by the Singapore Ministry of Education Tier 1 Grant No. RG 127/14 and the Polish National Agency for Academic Exchange NAWA Project No. PPN/PPO/2018/1/00007/U/00001.

\section*{Author contributions statement}
All authors researched and wrote this paper.

\section*{Additional information}
Processed data from tracking software is available on repository at: osf.io/zk9d6.
\\
\\
\textbf{Competing Interests:} The authors declare no competing interests.
\bibliography{cockroachbib}

\end{document}